# Rapid Prediction of Earthquake Ground Shaking Intensity Using Raw Waveform Data and a Convolutional Neural Network

Dario Jozinović[1,2], Anthony Lomax[3], Ivan Štajduhar[4], Alberto Michelini[1]

[1] *Istituto Nazionale di Geofisica e Vulcanologia, Via di Vigna Murata 605, 00143 Rome, Italy*
[2] *Department of Science, Università degli Studi Roma Tre, Via Ostiense, 159, 00154 Rome, Italy*
[3] *ALomax Scientific, 320 Chemin des Indes, 06370 Mouans-Sartoux, France*
[4] *Department of Computer Engineering, Faculty of Engineering, University of Rijeka, 51000 Rijeka, Croatia*

# Summary

This study describes a deep convolutional neural network (CNN) based technique for the prediction of intensity measurements (IMs) of ground shaking. The input data to the CNN model consists of multistation 3C broadband and accelerometric waveforms recorded during the 2016 Central Italy earthquake sequence for M ≥ 3.0. We find that the CNN is capable of predicting accurately the IMs at stations far from the epicenter and that have not yet recorded the maximum ground shaking when using a 10 s window starting at the earthquake origin time. The CNN IM predictions do not require previous knowledge of the earthquake source (location and magnitude). Comparison between the CNN model predictions and the predictions obtained with Bindi et al. (2011) GMPE (which require location and magnitude) has shown that the CNN model features similar error variance but smaller bias. Although the technique is not strictly designed for earthquake early warning, we found that it can provide useful estimates of ground motions within 15-20 sec after earthquake origin time depending on various setup elements (e.g., times for data transmission, computation, latencies). The technique has been tested on raw data without any initial data pre-selection in order to closely replicate real-time data streaming. When noise examples were included with the earthquake data, the CNN was found to be stable predicting accurately the ground shaking intensity corresponding to the noise amplitude.

## 1 Introduction

Rapid assessment of earthquake generated ground motions is a fundamental task of earthquake monitoring to provide information crucial to disaster response and public information. In recent years, rapid analysis has become of primary importance since advances in communication technology and the advent of social media have enabled public dissemination of (near) real-time information on events and their associated impact. To meet this challenge and to mitigate the impact of earthquakes, the seismological community has developed a number of earthquake early warning (EEW) systems (e.g., see Allen et al., 2009 and Satriano et al., 2011 for reviews; Kohler et al., 2017; Minson et al., 2018). These EEW systems are designed to detect earthquakes very rapidly (in a few seconds) and to provide early warning on the impending ground motion at selected target points.

On a longer timeline than conventional EEW, the ShakeMap software (Wald et al., 1999) was developed with the primary purpose of providing accurate maps of strong ground motion. These maps, typically available within 5-10 min after an earthquake, allow disaster risk managers (DRMs) to make preliminary assessments of the shaking impact. ShakeMap generates maps of ground shaking using earthquake source parameters (location, magnitude and the finite fault if available), intensity measurements (PGA, PGV, …), ground motion models (GMMs) and Vs30 maps as proxy to account for site amplifications. Ground motion interpolation is performed at points of the map which do not have ground motions recorded (e.g., Worden et al., 2018). The ShakeMap software has not been designed for EEW -- typically shakemaps become available only when the first location and magnitude estimation are available.  However, the ShakeAlert system for EEW in the western US has been recently upgraded to include the determination of ground motion using the same region-specific ground-motion prediction equations that are used by ShakeMap implementations in California, Oregon, and Washington (Given et al., 2018). In general, the time required to produce the first shakemaps depends on several factors such as data availability, and transmission and processing latencies, and the map accuracy depends on the density of the stations and on the quality of the data available (e.g., for the 2016 Central Italy M=6 August 24 mainshock, the first ShakeMaps were provided 6 minutes after origin time; Faenza et al., 2016).

In this study, in an effort to quantify as quickly as possible the level of ground shaking in a given area, we aim at predicting very rapidly the ground shaking intensity at a predefined set of seismic stations within a given seismic area by using a machine learning (ML) approach. The idea is to use waveforms up until only a few  seconds after origin time when only a few nearby stations have recorded the first P-waves to predict the ground shaking intensity at more distant stations.  The proposed technique does not require knowledge of the source parameters and it utilizes only a training set of earthquake waveforms recorded at a pre-configured network of recording stations. Implementation of the technique could be valuable when seeking a few seconds of warning to issue alerts for critical infrastructure points for potential failure such as highway or high velocity railways bridges, gas-pipelines,  industrial plants handling high risk and polluting chemicals, hospitals or schools.

We use the information contained in the recorded waveforms by adopting the convolutional neural network (CNN) model (Krizhevsky et al., 2012) to predict intensity measurements (IMs) at distant stations using only the initial N seconds after the origin time of the recordings at nearby stations. This implies that, depending on the value of N, the source receiver-geometry, and the duration of strong motion, a variable number of stations near the epicenter will contain also the maximum IM recorded for the target earthquake, whereas other more distant stations will not (Fig 4). The assumption is that the CNN model will be able to learn from the patterns of signal and noise across the input vector (i.e., relative amplitudes of the waveform signal in the window), and be able to make predictions on the maximum level of ground motion at the farther stations that have not yet recorded the peak amplitude values. Though this approach may not compete with the rapidity of an EEW system, it is expected to provide predictions of the ground motions similar or better to those obtained using EEW techniques and possibly the very first shakemaps, but at an earlier time.

While ML techniques have been used in seismology for over two decades (e.g., Chiaruttini et al., 1989; Dysart and Pulli, 1990), they have become more intensively used only in recent years (for more details see the reviews by Kong et al., 2018 and Bergen et al., 2018). ML has been applied to numerous problems: fast magnitude determination (Ochoa et al., 2018), ground motion prediction from source parameters (Derras et al., 2014; Alavi and Gandomi, 2011), earthquake detection (Reynen and Audet, 2017), insight into physics of labquakes (Hulbert et al., 2019), and many other seismological applications. A specific type of ML modelling technique, the CNN, has been used widely for waveform analysis. A CNN is a type of deep artificial neural network which uses convolutions as the fundamental building block for learning proper feature extraction, coupled with the potential for modelling highly-nonlinear classification/regression interactions between the variables. Perol et al. (2018) used a CNN to detect and locate earthquakes, by working directly on the seismic waveforms, without feature extraction. Similarly, Lomax et al. (2019) used a CNN for single-station, earthquake location, magnitude and depth determination from local to teleseismic scale lengths. Zhu and Beroza (2018) used a CNN to determine P and S wave arrival picks, also by analysing single-station seismic waveforms. Kriegerowski et al. (2018) created an earthquake location algorithm, by applying a CNN for analysing multi-station (10 stations) waveforms of clustered earthquakes. The successful applications of CNNs cited here, and the problem analysed in Kriegerowski et al. (2018), which is similar to the problem we address (analysis of multi-station waveforms), shows that CNN modelling is a viable approach for analysing multi-station waveforms towards rapid and accurate IM predictions.

## 2 Data

The input data used in this study consist of 3-component earthquake waveforms data from 2016 Central Italy sequence (Chiaraluce et al., 2017) recorded by 39 stations in the epicentral area and its surroundings. The dataset has been selected because of the dense network of stations in the area (Fig. 3), the large number of earthquakes and the importance of the sequence from the seismological perspective. We use earthquakes in the study area bounded by latitude [42°, 43.75°] and longitude [12.3°, 14°] which occurred from 1 January 2016 to 29 November 2016. All the events occur within crustal depths in the range 1.6km < D < 28.9km. Using these criteria, 915 earthquakes with magnitude M≥3 have been used (Figs 1 and 2). In the same study area there are 86 recording stations from the networks IV and XO available. In this study, we have selected a subset of 39 stations, which had at least 700 earthquakes recorded on the station (Fig. 3). The selected stations were all belonging to the IV network. Together with the event data, we have also selected 1037 examples of noise data, recorded from 30th August to 30th September 2016 recorded at all the selected stations. The criterion adopted for selecting the noise data consisted of windowing the continuous waveforms with a start time at least 180 seconds after and 20 seconds before the origin time of any earthquake in the area. The waveform data were downloaded using the INGV FDSN web services for HN* (acceleration) and HH* and EH* (velocity) channels, where * ∈ [E,N,Z]. The data were processed to remove the instrument response. Velocity data were differentiated to acceleration. When necessary, the data have been resampled to 100 Hz. For M < 4 earthquakes, the HH and EH channels were used after differentiation, and for earthquakes with M ≥ 4.0 the HN channels were used. This criterion avoided possible signal saturation and the use of true acceleration recordings for the larger magnitude earthquakes. If the velocimeter data were not available at one station with co-located accelerometer, the latter recording was used. For certain stations and for some earthquakes, the waveform data were completely missing and it was chosen to fill them with zeros. To this regard, there are different ways of filling missing data in ML (Garca-Laencina et al., 2010) and we chose to adopt zeroes as a natural way of applying the station dropout technique with the expectation of improving the generalisation properties of the model (similar to Kriegerowski et al., 2018). Our data set includes stations with missing earthquake recordings at random only after the Amatrice August 24th M6.0 main shock. This is because 125 earthquakes that had occurred before the Amatrice earthquake were not recorded by the temporary stations installed immediately after this main shock.

All data have been extracted using a 50 s window starting at origin time. We use origin time as a convenience for aligning the waveforms. In practice, however, any reference time before the first recorded P wave at the first recording station could be used instead. Data that started more than 0.1 seconds (178 cases) after origin time have been manually inspected, and the faulty ones (instrument malfunctioning; 24 cases) were removed. Those that were retained (154 cases) were prepended with zeros between origin time and the start of data.

The *target data* consisted of the IMs associated to each recording: peak ground acceleration (PGA), peak ground velocity (PGV), spectral acceleration (SA) at 0.3 s, 1 s and 3 s periods. For the stations that had no data, the IMs were calculated using USGS ShakeMap software with the latest configuration for Italy (Michelini et al., 2019). ShakeMaps have been calculated using the IMs from all available stations. This has been done to ensure no missing output data (target variables), and it is expected to be beneficial in the learning stage of the ML procedure. This approach, however, can introduce some error since the ShakeMap prediction may be incorrect given the uncertainties in the estimation of the ground motion. We note, however, that Michelini et al. (2019) have shown that nearly no systematic bias in the prediction of the maximum IMs was observed with the new ShakeMap configuration giving us good confidence that the values inserted as target values are statistically relevant. The target data for the noise waveforms were the maximum amplitudes recorded at the station inside the window used as input. In the case of missing inputs in the noise waveforms, the target has been set to zero.

## 3 Method and training

The ML model adopts a CNN developed using Keras (see Data and Resources). Input to the model is a combination of all the waveform data (all 39 stations) for a given earthquake, for an input array size (39, N, 3), where 39 is the number of stations, N is the number of samples (with sampling rate of 100 samples per second) and 3 is the number of components. The ordering of the stations is always preserved. All the waveform data for each earthquake start at the event origin time. The data are normalised by the input maximum (i.e., the largest amplitude observed across all stations within the time window), and this maximum is saved as the normalisation value which is later inserted into the network. The outputs are arrays of size (39,5), where 39 is the number of stations, and 5 is the number of predicted IMs per station. The base-10 logarithm has been applied to all the IMs (i.e., $\log_{10}$IM).

The architecture is based on Kriegerowski et al. (2018) with minor modifications (Fig. 5). The network consists of 3 convolutional layers and one fully connected layer. Input to the model is a combination of all the data (all 39 stations) for a given earthquake, for an input array size (39, N, 3), where 39 is the number of stations used, N is the temporal length in samples (with sampling rate of 100 samples per second) and 3 is the number of components used. The first two convolutional layers, having filters of size 32 and 64, respectively, have kernels with height of 1 and width of 125, whose purpose is to learn the temporal patterns station by station. The third convolutional layer, having 64 channels and kernels with height 39 and width 5, gathers the cross station information. The first two convolutional layers have the stride (1,2), and the third layer has the stride (39,5). The last convolutional layer is then flattened and concatenated with the normalization value of the input, and then fed to a fully connected layer. Finally, the last fully connected layer produces an array of size (39,5) with continuous values inside. The ReLU activation function, a=max(0,x), has been used in all layers, except the last layer in which linear activation has been used. L2 regularization has been applied to the convolutional layers with regularization rate of $\lambda=10^{-4}$. Furthermore, 40% dropout rate has been applied before the last fully connected layer (Goodfellow et al., 2016). The data had been split randomly into training (80%) and test (20%) subsets. For evaluation of the performance of the CNN model 5-fold cross-validation has been used, which splits the randomly permuted dataset into 5 equally-sized disjunct subsets, and uses each of them as the test set to the model trained on the remaining 4 subsets. The batch size used for optimisation was 5, and the mean squared error (MSE) function was used as the loss function in the model. Hyperparameter values for model optimization were based on Kriegerowski et al. (2018) with minor modifications. The model was trained for 12 epochs (the training history plot is shown in Fig. 5 of the electronic supplement). The training took approximately 3 minutes on an Nvidia GTX 1060 6GB for a 10 second input window.

## 4 Results

### 4.1 Window length

In the first part of this study, we explored the window length after origin time needed to make reliable predictions of the maximum values of the IMs. We used window lengths of 7 s, 10 s and 15 s. The results are shown only for PGA, as other IMs follow similar trends. For the 7 s window, the MSE of the residuals between the base-10 logarithms of observed and predicted values (i.e., $log_{10}(IM_{true}/IM_{predicted})$) was 0.228. For the 10 s window, the MSE is reduced to 0.176, and for the 15 s window, the misfit was further reduced to 0.165. A significant drop in performance is thus occurring for shorter windows.

The vertical stride of the observed values at approximately -0.9 derived from the occasional bad input waveforms (examples in the Figs 1 and 2 in the supplement). In what follows, we adopt a 10 s window as a good compromise between accuracy of the predictions and the timeliness.

### 4.2 Model performance

The performance of the CNN model has been evaluated using 5-fold cross-validation. The results from all 5 test sets were then averaged, and are presented next. The target values are the recorded IMs (93%), and the ShakeMap predictions of IMs that were used when no recordings were available (7%). The residuals ($log_{10}(IM_{obs}/IM_{predicted})$) have been calculated together with their mean, median and standard deviation. Large residual values ($log_{10}(IM_{obs}/IM_{predicted}) >|1|$) were removed resulting into 87.49% of the data kept for the ShakeMap predictions and 92.00% for the observed IMs. Table 1 and Fig. 7 show the performance of the CNN model on the observed values, along with the ShakeMap predictions.

As noted above, the ShakeMap predictions are used for the training set when no observed data are available, and the input is set to a zero value time series. To this regard, we find it notable the capacity of the adopted CNN algorithm to learn how to treat those stations that occasionally may be missing data and the target value consists of the ShakeMap predictions. In general, a slight positive bias in the CNN predictions of IM values is observed, indicating that the CNN model is slightly underpredicting the observed values with the exception of the ShakeMap predicted SA(3.0).

To test the CNN model performance against a baseline case, we compare our results to the predictions obtained using the GMPE by Bindi et al. (2011) calibrated for Italy. For the IM predictions using Bindi et al. (2011), we used the INGV catalog magnitude and location (see data and resources) and the IM predictions are corrected appropriately using the station EC8 site classes. No between-event correction (Al Atik et al., 2010), however, has been applied when predicting the ground motion using the GMPE by Bindi et al. (2011). The outliers have been discarded using the same criteria used earlier which left us with 90.3% of the data. For the CNN model performance, only the results obtained for the observed IM values are significant. The residuals $log_{10}(IM_{observed}/IM_{predicted})$ have been calculated for the results obtained with the GMPE model, in the same way as for the CNN model.

The mean, median and standard deviation for the CNN and the GMPE models are reported in Table 1 and they are shown in Fig. 7. We find that the median values (of the differences between the logarithms of observed and predicted values), which are an expression of the model bias, are significantly reduced in the CNN model, especially for PGV, PSA(0.3) and PSA(1.0), yet they are slightly higher for PGA. Concerning the standard deviation, we find that the values are comparable for the two models, and are in agreement with those reported by Bindi et al. (2011) who reported a standard deviation between 0.34 and 0.38 in $log_{10}$ unit.

When applied on the 3 largest events in the sequence (using 5-fold cross-validation), with magnitudes of M=[5.9, 6.0, 6.5], the CNN model performance deteriorates (Fig. 8), compared to the model performance on events with smaller magnitude. These results are due to the unbalanced dataset with few training examples at large magnitude (Fig. 2), which makes learning at larger magnitudes difficult. More specifically, the poor performance for the *M 6.0* event can be also attributed to the use of only 24 station waveforms in the input data, as no temporary stations had yet been installed in the area before the August 24, 2016, *M 6.0* main event. This same reasoning, however, does not hold for the performance differences between the *M 6.5* and *M 5.9* events (33 and 35 stations, respectively) with the former one performing much more poorly than the latter. One possible and partial explanation for this performance disparity could be ascribed to the difference in the pattern of the input waveforms (Figs 3 and 4 in the Supplement). Since the *M 5.9* event is located to the north where there is a larger number of stations, this event features more stations with "significant" earthquake P-wave (or even S-wave for the nearest ones) signal information within the 10 s window when compared to the *M 6.5* earthquake. More in general terms, the rather poor predictive performance observed for the *M 6.0* and *M 6.5* earthquakes derives primarily from the lack of earthquake waveform data at the larger magnitudes.

### 4.3. Including noise examples in the data

It is important to evaluate the CNN model performance in the presence of noise-only data, in which case the predicted IM's should reflect the maximum noise level at each station. We added 1037 examples of the noise data to both the training and the test subset, and again we used the maximum PGA of the input waveforms as the training and test target variables. The results (Fig. 9) show that the CNN model is also able to predict the target IMs of the noise data reasonably well. The target data for the non-existing noise waveforms (for which inputs are filled with zeros) are zero, and they are not shown in the plot.

## 5 Discussion

In this study, we have shown that a CNN model can be used to predict accurately earthquake IMs at recording stations using only raw, multi-station waveforms with a 10 s time window starting at the earthquake origin time. That is, from the very first recordings at stations near the epicenter, it is possible to predict accurately the IMs at farther stations that have not yet recorded the earthquake signal or its maximum values. To this end, we exploit the 3C station data (i.e., waveforms) and the pattern of the waveforms of the recording network without any knowledge concerning the earthquake location or magnitude.

Input data consist of the recorded acceleration waveforms, while the target variables are PGA,

PGV and SA at 0.3 s, 1 s and 3 s periods. The CNN generally performs equally well regardless of the IM type. As expected, model performance improves when longer time windows (e.g., 15 s time windows) are used as input.

Because of the Gutenberg-Richter distribution of earthquake magnitudes, our dataset is severely unbalanced with a much larger frequency of smaller events than larger ones. As shown in Fig. 8, this makes it difficult to predict the IMs of larger events. A possible solution to this problem could follow from the use of data augmentation (e.g., Chollet, 2018) by introducing more training data for the larger magnitudes after applying random transformations to the existing larger events. Another alternative can consist of calculating synthetic seismograms at the same recording stations for additional (hypothetical) larger events in the same area. Another approach towards mitigating problems due to small datasets is the use of transfer learning methods (Pan and Yang (2010)). These methods allow the CNN model to subspecialize by training on a smaller dataset, using a pre-trained model which was previously trained on a larger dataset. Following Chollet (2018), we could even use a training network developed elsewhere and for different purposes, but with a very large dataset, and use it to our goal of IM prediction. Similarly, the network developed in this study could be adopted as a pre-trained model for other areas where IM predictions are required.

To test the methodology, we have used raw waveforms without any preliminary data cleaning, in order to simulate real-time analysis where missing or erroneous data due to equipment malfunctioning or data transmission problems is common. To compensate for these deficiencies, we found that replacing missing data with zero values and the adoption of ShakeMap predicted target data values is suitable to ensure that the target data exist for all cases during learning. We have shown that by adopting this strategy, the CNN model is still capable of predicting the IMs fairly accurately. In practice, this is somewhat similar to applying the station dropout technique (Kriegerowski et al., 2018) to the training procedure. We have found that the CNN model was able to accurately predict the ground motions at the stations with missing data, albeit with a slightly lower accuracy when compared to the IMs of the stations which have no missing data. This all suggests that the CNN model is able to learn the seismic wave propagation characteristics and can compensate for the presence of missing data when using imperfect target data, like the ShakeMap predictions, for learning.

To verify the predictive performance of our CNN model, and compare it against the results of a more conventional way of predicting the IMs, we have used the GMPEs proposed by Bindi et al. (2011) which are adopted in the configuration of the ShakeMap for Italy (Michelini et al., 2019) to predict the IMs at our selected stations. The results shown in Fig 7 indicate the CNN model to be superior in terms of IM residuals. This follows from i.) the CNN model being independent of earthquake magnitude uncertainties (i.e., no need for between-event correction terms), ii.) our source-receiver geometry featuring similar wave paths within a rather small area, and all earthquakes occurring approximately in the same source region. and iii.) the ability of the CNN model to capture local site anomalies and compensate for them. We note that these points are all taken into account by CNN modelling procedure automatically by learning directly from the 3C waveforms.

As noted above, the earthquakes used in these experiments are concentrated spatially which significantly reduces the variability at a given station resulting from different paths, and this may be one of the main reasons for the good quality of the results obtained. We are confident, however, that by using a bigger dataset, with different spatial distribution of epicenters, the CNN model will likely be able to learn the path and site characteristics of each station and for each zone where the earthquakes occur. This aspect will be addressed in future studies encompassing larger areas and datasets.

Our results with earthquake and noise only data (Fig. 9), have shown that the methodology is capable to predict the ground motion adequately. This fact combined with the translational invariance characteristics of convolutional neural networks suggests that the technique could be used in a real-time setup with data streamed continuously to predict the IMs seamlessly at farther stations as the earthquake nearest stations start recording the first earthquake generated signals.

## 6 Conclusions

A CNN model has been used to predict IMs (PGA; PGV; PSA 0.3s, 1s, 3s) with satisfying accuracy adopting multi-station, 3C 10s window waveforms starting at origin time. The waveforms come from a data set recorded by 39 stations from a set of 915 earthquakes with M ≥ 3.0 and 1037 only noise examples, for a total of 33855 earthquake waveforms, and 40443 noise waveforms.

The IM predictions do not require previous knowledge of earthquake location, distance and magnitude. When using longer time windows the performance of the CNN improved but the selected 10 s window of our setup appears to be a good compromise between accuracy and timeliness. No feature extraction or preprocessing was to be applied on the data because CNN models do it automatically. The performance of the CNN model does not depend on the IM type. We have found that the CNN model was able to predict with satisfying accuracy the IMs for stations which had no input data when training. Comparison between the CNN results and those obtained with the Bindi et al. (2011) ground motion model, calibrated with earthquakes in Italy, has shown that the CNN model does not suffer from prediction bias while featuring similar variance of the residuals.

The performance of the CNN model degrades for large magnitudes (M>5) because of the small number of earthquakes (5) available for training. We have also observed that the CNN model performs well when noise only data were included in the dataset. This results gives us good confidence that the proposed analysis could be implemented in a real-time analysis configuration. Although the technique is not strictly designed for earthquake early warning, we found that it could provide useful estimates of ground motions within 15-20 sec after earthquake origin time depending on the data transmission infrastructure, latencies and processing time requirements.

The study uses a specific set of stations and earthquakes concentrated spatially and it could be applied to larger areas with more widespread seismicity as long as enough data are available for training. The main purpose of this study was to show that the technique can provide quite satisfactory results in terms of predicted ground motion only by learning patterns of multistation 3C waveforms.

## 7 Data and Resources

Earthquake catalogue and waveform data have been downloaded through the INGV FDSN web services (http://terremoti.ingv.it/en/webservices_and_software). Waveforms have been downloaded and processed using python library Obspy (Beyreuther et al. (2010) ). IMs for the stations with no data have been calculated using USGS ShakeMap 4 (http://usgs.github.io/shakemap/sm4_index.html). The CNN model has been developed using the Keras Python Deep Learning library (Chollet et al. (2015)). The code and the data of this paper will be available on https://github.com/djozinovi/CNNpredIM.

# 8 Acknowledgements

The research has been funded by SERA EU project (Seismology and Earthquake Engineering Research Infrastructure Alliance for Europe;contract n. 730900). It has also greatly benefited from a travel grant for a short term scientific mission provided by G2NET, COST action CA17137. The authors would like to thank the operators of RSN (INGV) and the USGS ShakeMap team of developers.

## 8.1 Author contribution statement

The topic of this study was conceived by Alberto Michelini, with contributions from Anthony Lomax and Dario Jozinović. Dario Jozinović prepared, downloaded and processed the earthquake data; Alberto Michelini prepared and processed the ShakeMap data. The CNN model has been set up and trained by Dario Jozinović with contribution from Ivan Štajduhar. The results have been analysed by Dario Jozinović with contribution by Alberto Michelini. The final manuscript has been written by Dario Jozinović and Alberto Michelini, with contributions by Anthony Lomax and Ivan Štajduhar.

# References list

Alavi, A. H., and Gandomi, A. H. (2011). "Prediction of principal ground-motion parameters using a hybrid method coupling artificial neural networks and simulated annealing". *Computers & Structures*, *89*(23-24), 2176-2194.

Allen, R. M., Gasparini, P., Kamigaichi, O., and Bose, M. (2009). "The Status of Earthquake Early Warning around the World: An Introductory Overview". *Seismological Research Letters*, 80(5), 682–693. http://doi.org/10.1785/gssrl.80.5.682

Atik, L. A., Abrahamson, N., Bommer, J. J., Scherbaum, F., Cotton, F., and Kuehn, N. (2010). "The Variability of Ground-Motion Prediction Models and Its Components". *Seismological Research Letters*, 81(5), 794–801. http://doi.org/10.1785/gssrl.81.5.794

Bergen, K. J., Johnson, P. A., Maarten, V., and Beroza, G. C. "Machine learning for data-driven discovery in solid Earth geoscience." *Science* 363.6433 (2019): eaau0323.

Beyreuther, M., Barsch, R., Krischer, L., Megies, T., Behr, Y., and Wassermann, J. (2010). "ObsPy: A Python toolbox for seismology." *Seismological Research Letters* 81.3 (2010): 530-533.

Bindi, D., Pacor, F., Luzi, L., Puglia, R., Massa, M., Ameri, G., and Paolucci, R. (2011). "Ground motion prediction equations derived from the Italian strong motion database." *Bulletin of Earthquake Engineering* 9.6 (2011): 1899-1920.

Chiaraluce, L., Di Stefano, R., Tinti, E., Scognamiglio, L., Michele, M., Casarotti, E., Cattaneo, M., De Gori, P., Chiarabba, C., Monachesi, G. and Lombardi, A. "The 2016 central Italy seismic sequence: A first look at the mainshocks, aftershocks, and source models." *Seismological Research Letters* 88.3 (2017): 757-771.

Chollet, F. (2018). "Deep learning with Python". New York, NY: Manning Publications.

Chollet, François and others, "Keras", 2015 Keras

Chiaruttini, Claudio, V. Roberto, and F. Saitta. "Artificial intelligence techniques in seismic signal interpretation." *Geophysical Journal International* 98.2 (1989): 223-232.

Derras, Boumédiène, Pierre Yves Bard, and Fabrice Cotton. "Towards fully data driven ground-motion prediction models for Europe." *Bulletin of earthquake engineering* 12.1 (2014): 495-516.

Dysart, Paul S., and Jay J. Pulli. "Regional seismic event classification at the NORESS array: seismological measurements and the use of trained neural networks." *Bulletin of the Seismological Society of America* 80.6B (1990): 1910-1933.

Faenza, Licia, Valentino Lauciani, and Alberto Michelini. "The ShakeMaps of the Amatrice, M6, earthquake." *Annals of Geophysics* 59 (2016).

Garca-Laencina, Pedro J., Jos-Luis Sancho-Gómez, and Anbal R. Figueiras-Vidal. "Pattern classification with missing data: a review." *Neural Computing and Applications* 19.2 (2010): 263-282.

Given, D.D., Allen, R.M., Baltay, A.S., Bodin, P., Cochran, E.S., Creager, K., de Groot, R.M., Gee, L.S., Hauksson, E., Heaton, T.H., Hellweg, M., Murray, J.R., Thomas, V.I., Toomey, D., and Yelin, T.S. "Revised technical implementation plan for the ShakeAlert system—An earthquake early warning system for the West Coast of the United States: U.S. Geological Survey Open-File Report 2018–1155, 42 p." (2018). https://doi.org/10.3133/ofr20181155.

Goodfellow, I., Bengio, Y., & Courville, A. (2016). "Deep learning". *MIT press*.

Hulbert, C., Rouet-Leduc, B., Johnson, P. A., Ren, C. X., Rivière, J., Bolton, D. C., and Marone, C. "Similarity of fast and slow earthquakes illuminated by machine learning." *Nature Geoscience* 12.1 (2019): 69.

Iervolino, I., Giorgio, M., Galasso, C., and Manfredi, G. "Uncertainty in early warning predictions of engineering ground motion parameters: What really matters?." *Geophysical research letters* 36.5 (2009).

INGV Seismological Data Centre. (2006, January 1). Rete Sismica Nazionale (RSN). Istituto Nazionale di Geofisica e Vulcanologia (INGV), Italy. https://doi.org/10.13127/sd/x0fxnh7qfy

Istituto Nazionale Di Geofisica E Vulcanologia (INGV). (2017). Rete sismica del gruppo EMERSITO, sequenza sismica del 2016 in Italia Centrale [Data set]. Istituto Nazionale di Geofisica e Vulcanologia (INGV). https://doi.org/10.13127/sd/7txegdo5x8

Kohler, M.D., Cochran, E.S., Given, D., Guiwits, S., Neuhauser, D., Henson, I., Hartog, R., Bodin, P., Kress, V., Thompson, S. and Felizardo, C. "Earthquake early warning ShakeAlert system: West coast wide production prototype." *Seismological Research Letters* 89.1 (2017): 99-107.

Kong, Q., Trugman, D. T., Ross, Z. E., Bianco, M. J., Meade, B. J., and Gerstoft, P. "Machine learning in seismology: Turning data into insights." *Seismological Research Letters* 90.1 (2018): 3-14.

Kriegerowski, Marius, et al. "A deep convolutional neural network for localization of clustered earthquakes based on multistation full waveforms." *Seismological Research Letters* 90.2A (2018): 510-516.

Krizhevsky, A., Sutskever, I. and Hinton, G. E. "ImageNet Classification with Deep Convolutional Neural Networks." *Adv. Neural Inf. Process. Syst.* 25, 1097--1105 (2012).

Lomax, Anthony, Alberto Michelini, and Dario Jozinović. "An investigation of rapid earthquake characterization using single‑station waveforms and a convolutional neural network." *Seismological Research Letters* 90.2A (2019): 517-529.

Luzi, L., Puglia, R., Pacor, F., Gallipoli, M. R., Bindi, D., and Mucciarelli, M. "Proposal for a soil classification based on parameters alternative or complementary to Vs, 30." *Bulletin of Earthquake Engineering* 9.6 (2011): 1877-1898.

Michelini, A., Faenza, L., Lanzano, G., Lauciani, V., Jozinovic, D., Puglia, R., and Luzi, L. (2019). "The New ShakeMap in Italy: Progress and Advances in the Last 10 Yr." *Seismological Research Letters* 91.1 (2019): 317-333.

Minson, S. E., Meier, M. A., Baltay, A. S., Hanks, T. C., and Cochran, E. S. (2018). "The limits of earthquake early warning: Timeliness of ground motion estimates." *Science advances* 4.3 (2018): eaaq0504.

Ochoa, Luis H., Luis F. Niño, and Carlos A. Vargas. "Fast magnitude determination using a single seismological station record implementing machine learning techniques." *Geodesy and Geodynamics* 9.1 (2018): 34-41.

Pan, S. J. and Yang, Q. "A Survey on Transfer Learning". *IEEE Trans. Knowl. Data Eng*. 22, 1345–1359 (2010).

Perol, Thibaut, Michaël Gharbi, and Marine Denolle. "Convolutional neural network for earthquake detection and location." *Science Advances* 4.2 (2018): e1700578.

Reynen, Andrew, and Pascal Audet. "Supervised machine learning on a network scale: Application to seismic event classification and detection." *Geophysical Journal International* 210.3 (2017): 1394-1409.

Satriano, C., Wu, Y.-M., Zollo, A., and Kanamori, H. (2011). "Earthquake early warning: Concepts, methods and physical grounds." *Soil Dynamics and Earthquake Engineering* 31.2 (2011): 106-118.

Wald, D. J., Quitoriano, V., Heaton, T. H., Kanamori, H., Scrivner, C. W., and Worden, C. B. "TriNet "ShakeMaps": Rapid generation of peak ground motion and intensity maps for earthquakes in southern California." *Earthquake Spectra* 15.3 (1999): 537-555.

Worden, C. B., Thompson, E. M., Baker, J. W., Bradley, B. A., Luco, N., and Wald, D. J. (2018). "Spatial and spectral interpolation of ground‑motion intensity measure observations." *Bulletin of the Seismological Society of America* 108.2 (2018): 866-875.

Zhu, Weiqiang, and Gregory C. Beroza. "PhaseNet: a deep-neural-network-based seismic arrival-time picking method." *Geophysical Journal International* 216.1 (2018): 261-273.

# Tables

Table 1: IMs' residual statistics for the CNN predictions for the observed IMs (for the stations having recorded data) and, the ShakeMap predictions (for the stations that had no recorded data). The last three columns show the values obtained using the Bindi et al. (2011) GMPE to estimate the IMs.

| IM | Observed median | Observed mean | Observed STD | ShakeMap median | ShakeMap mean | ShakeMap STD | GMPE median | GMPE mean | GMPE STD |
|---|---|---|---|---|---|---|---|---|---|
| **PGA** | 0.038 | 0.035 | 0.346 | 0.059 | 0.038 | 0.372 | 0.013 | 0.017 | 0.352 |
| **PGV** | 0.036 | 0.034 | 0.338 | 0.043 | 0.041 | 0.380 | -0.174 | -0.151 | 0.33 |
| **SA(0.3)** | 0.031 | 0.031 | 0.34 | 0.056 | 0.046 | 0.37 | -0.284 | -0.252 | 0.359 |
| **SA(1.0)** | 0.029 | 0.034 | 0.338 | 0.001 | 0.017 | 0.374 | -0.207 | -0.198 | 0.303 |
| **SA(3.0)** | 0.019 | 0.027 | 0.374 | -0.037 | -0.012 | 0.404 | 0.026 | 0.083 | 0.368 |

# Figures

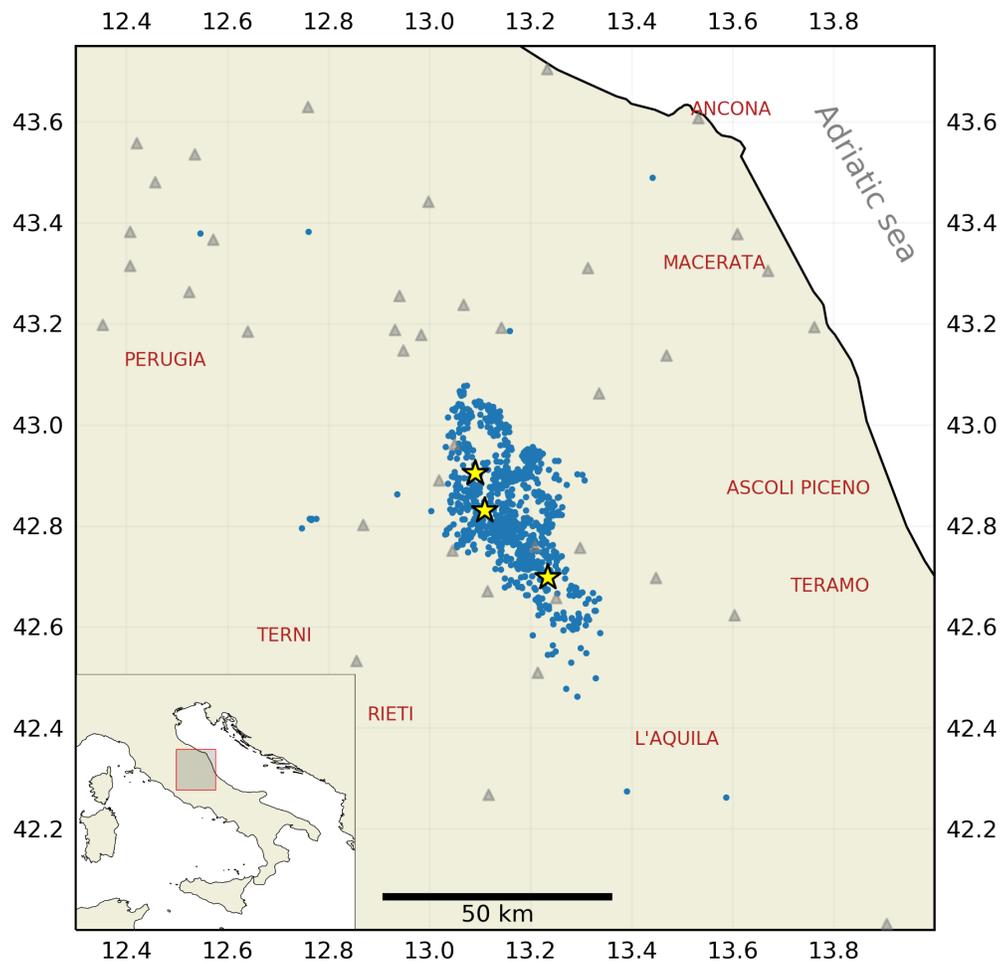

Figure 1. Blue points represent earthquake epicenter locations, stars represent the three largest events of the 2016 Central Italy sequence and the light gray triangles represent stations.

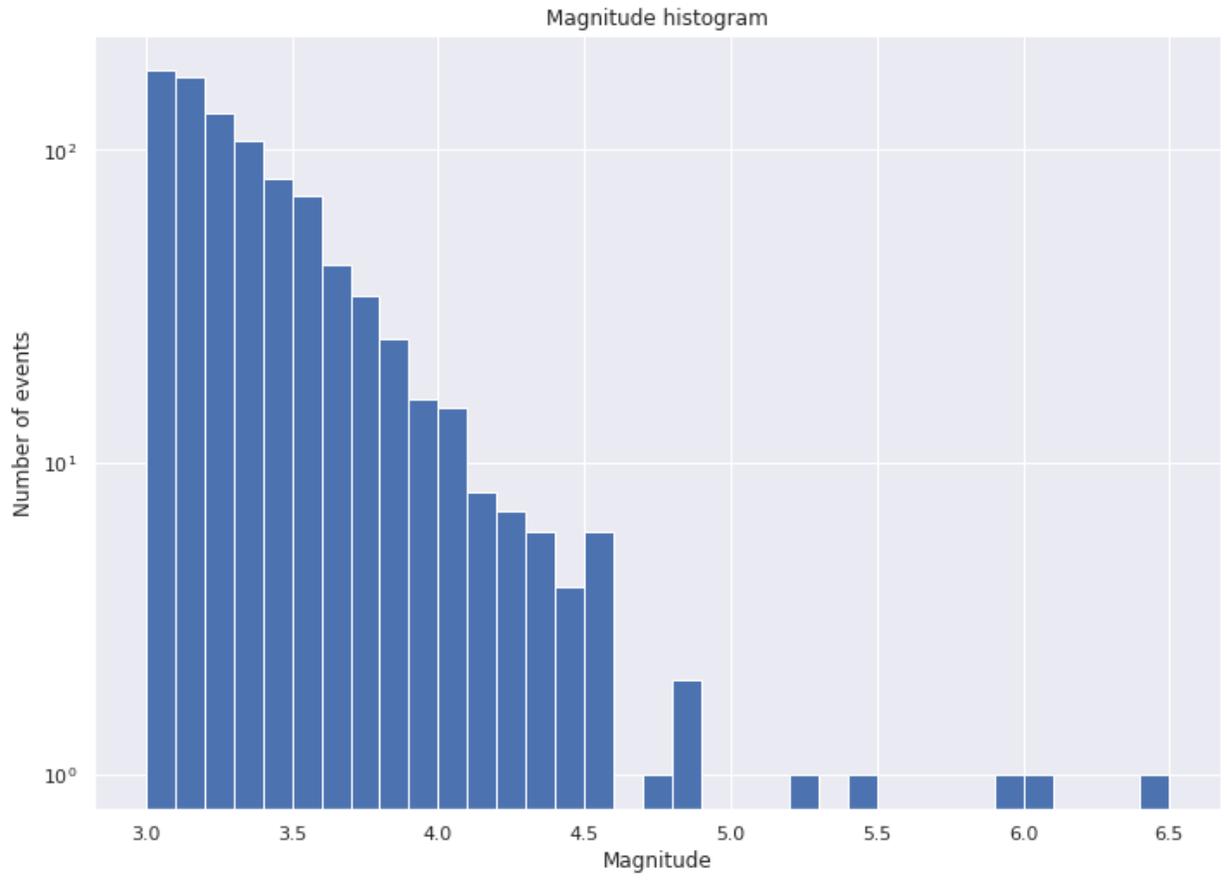

Figure 2. Earthquake-Magnitude distribution of the selected earthquakes.

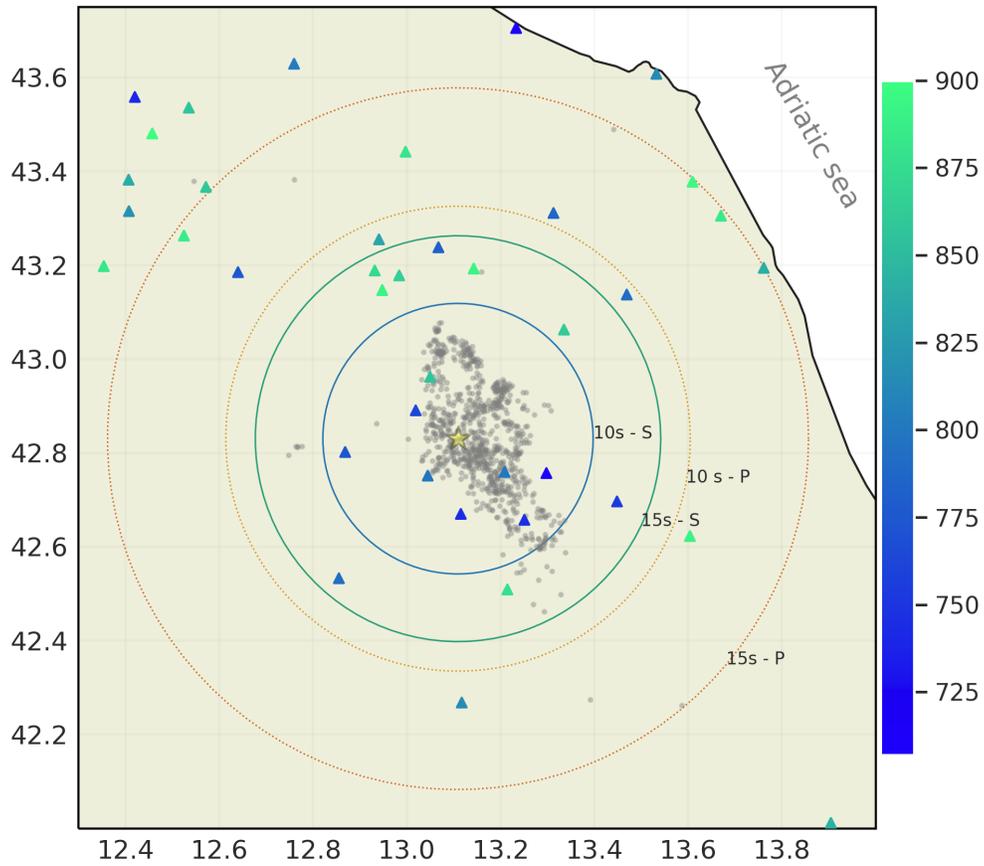

Figure 3. Stations used, with colors showing the number of earthquakes recorded per station. The circles represent the approximate travel times of direct P arrivals using Vp=5.5 and Vs=3.2 of the biggest event in sequence (M6.5 2016/10/30; shown with a star). Earthquakes as light gray points.

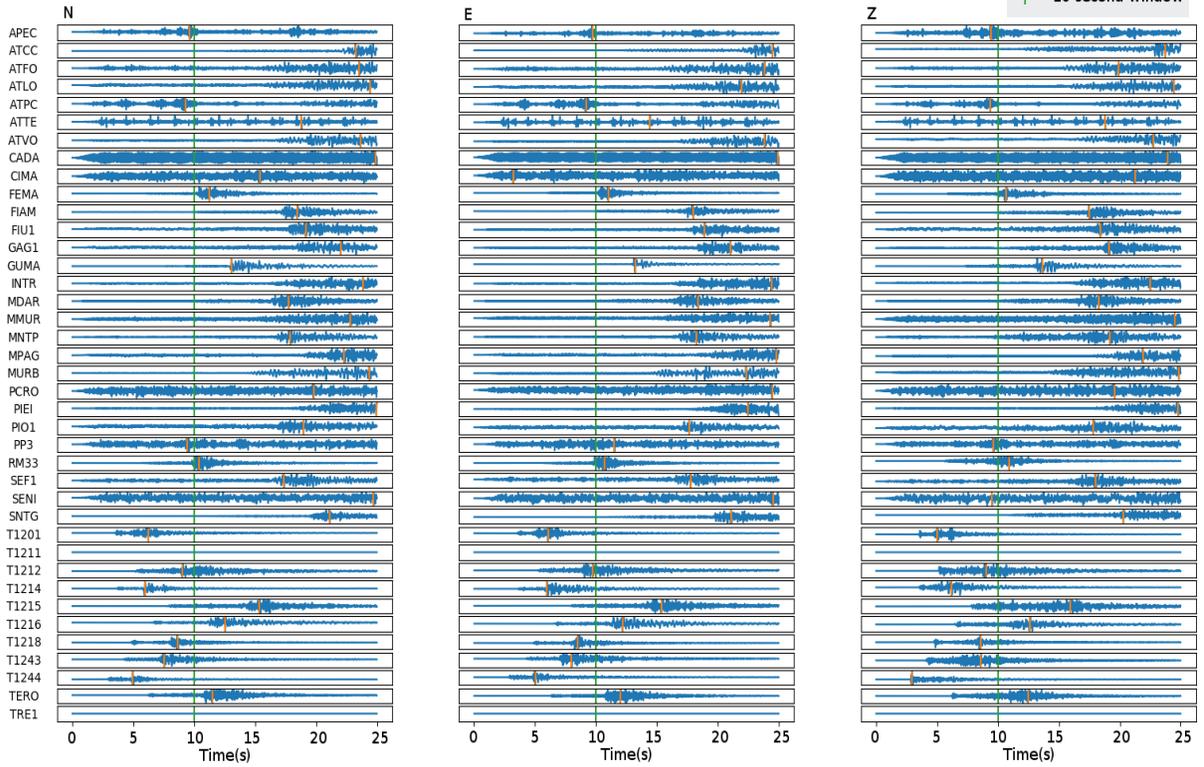

Figure 4. Non-normalised input example. Orange vertical lines show the trace peak value. Green vertical lines show the end of the 10 second window.

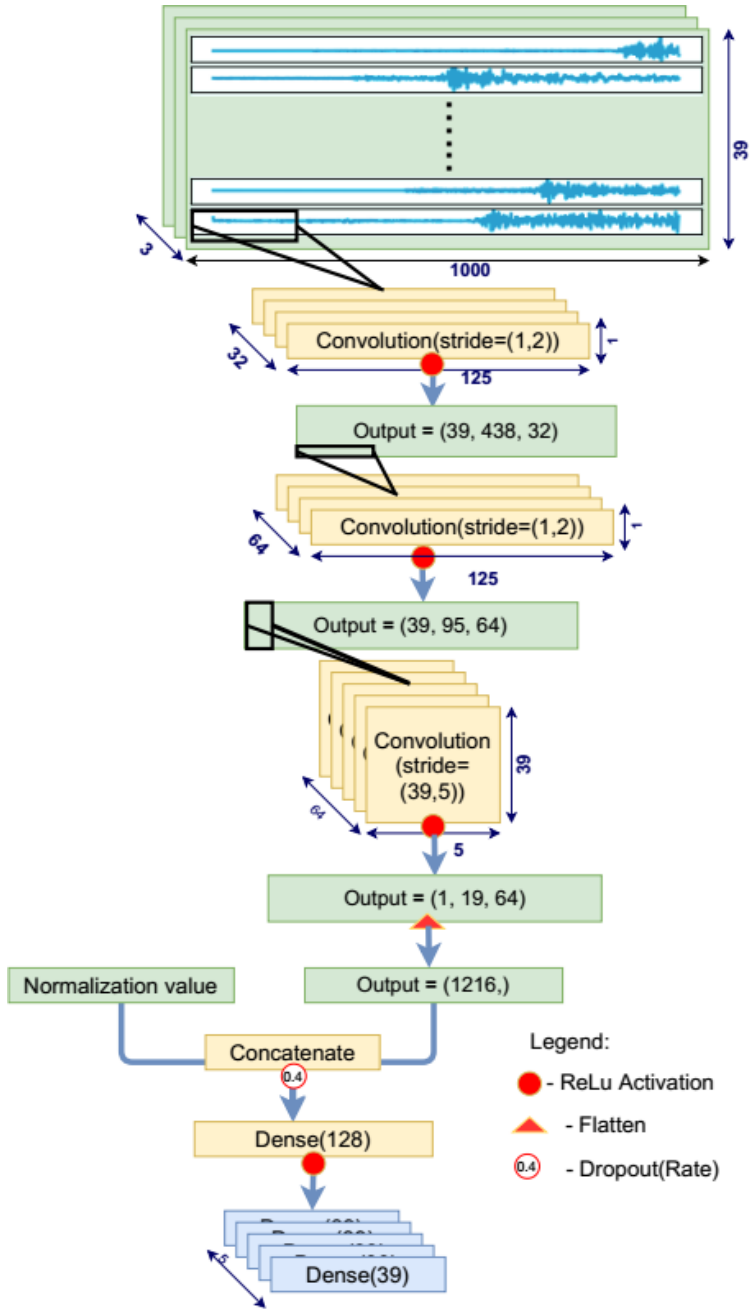

Figure 5. The architecture of the CNN model used. Boxes shaded yellow represent filter banks and operators, whereas boxes shaded green represent activations. Parenthesised vectors denote the dimensions of the outputs in question (height, width, depth).

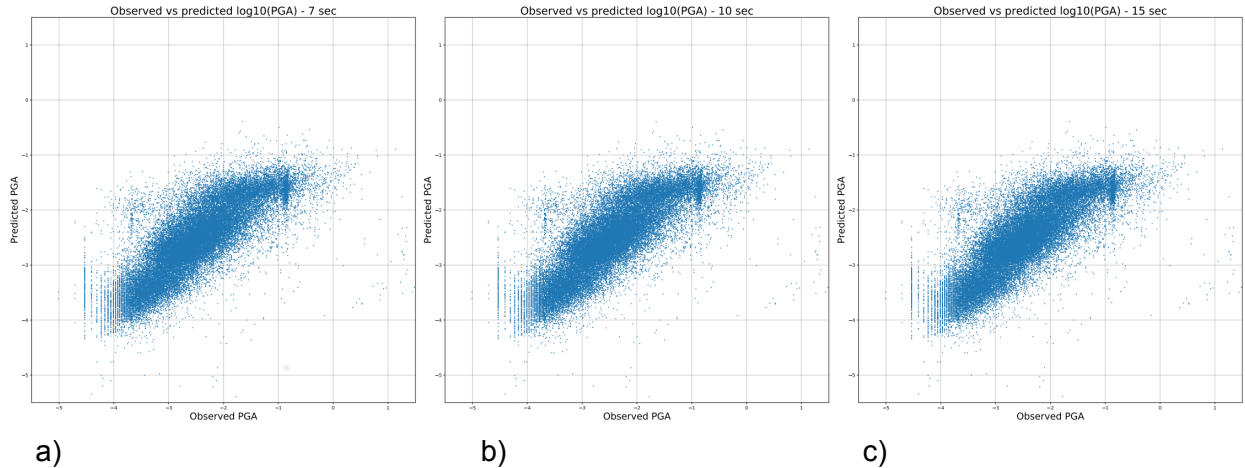

Figure 6. $\log_{10}(PGA_{observed})$ vs $\log_{10}(PGA_{predicted})$ for different time window lengths: a) 7 s length b) 10 s length c) 15 s length. The units are $\log(m/s^2)$.

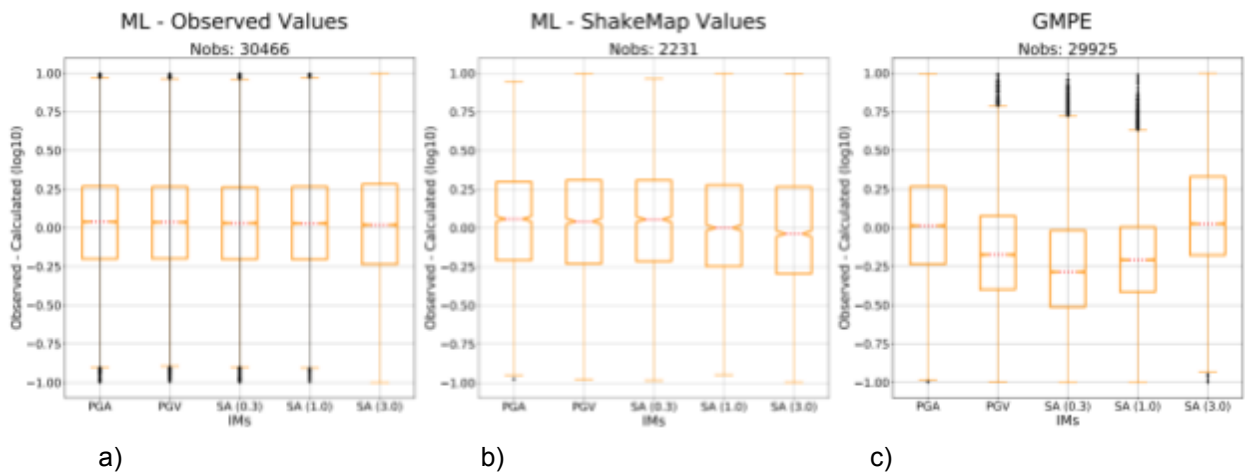

Fig 7. Boxplots for the residuals $IM_{observed} - IM_{predicted}$: a) CNN model results for the observed IMs, b) CNN model results for the ShakeMap predicted IMs, c) GMPE results.

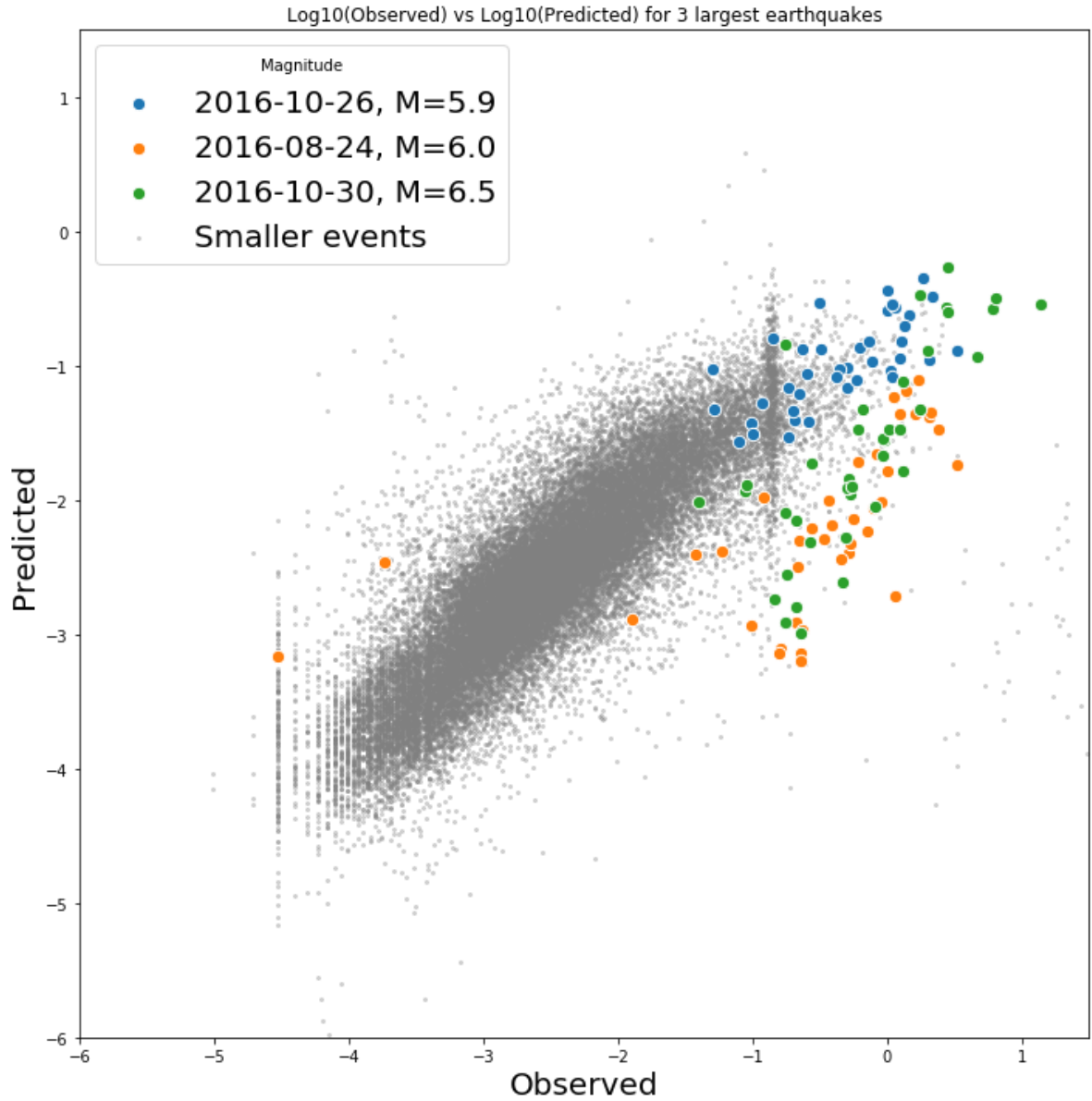

Figure 8: $\log_{10}(PGA_{observed})$ vs $\log_{10}(PGA_{predicted})$ for the three largest events in the sequence(using 10 s waveforms). The vertical stripe visible at approximately -0.9 $\log_{10}(PGA_{obseved})$ results from data logger equipment malfunctioning

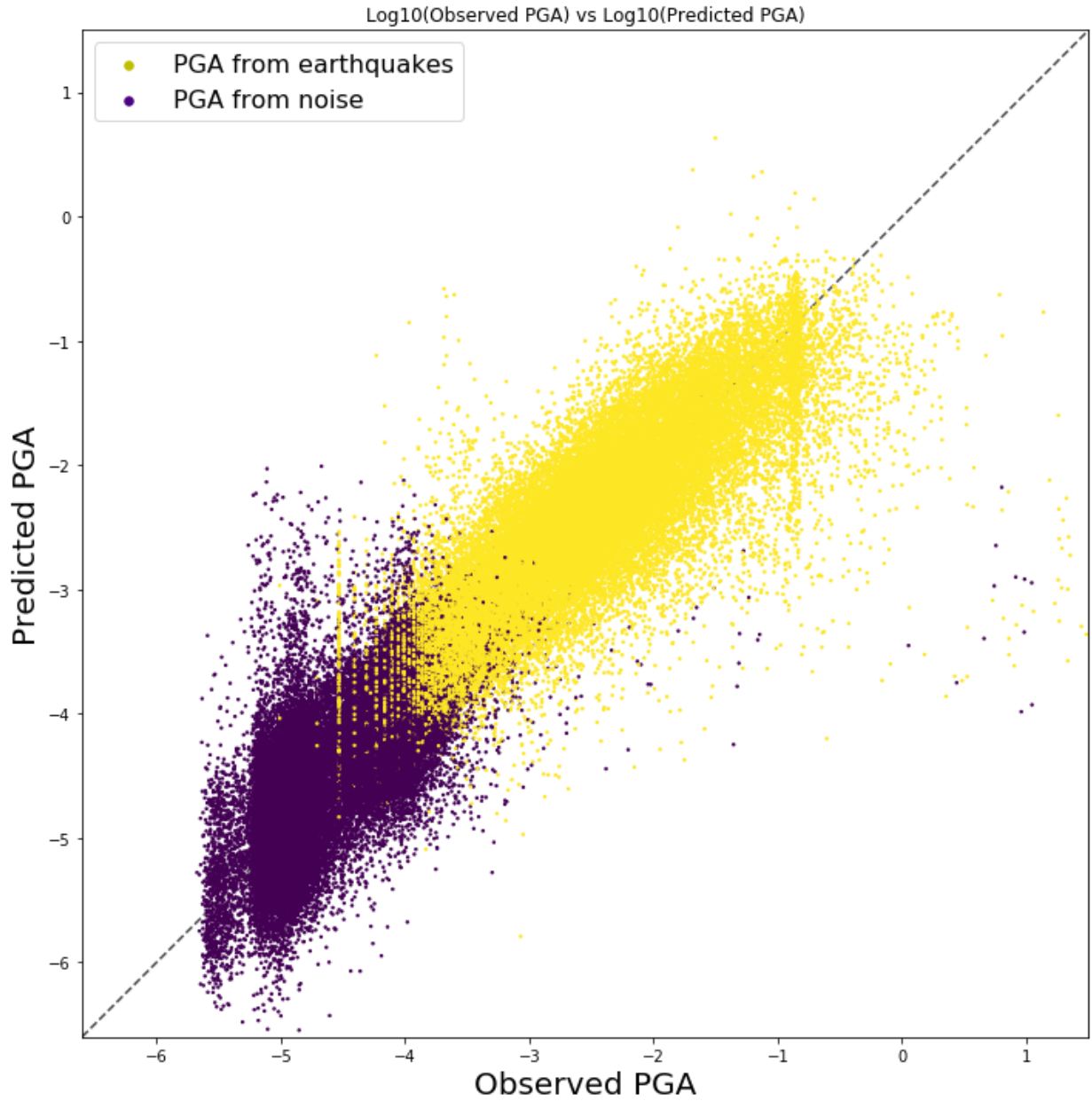

Figure 9. $\log_{10}(PGA_{obseved})$ vs $\log_{10}(PGA_{predicted})$ for training with noise and event data. The vertical stripe visible at approximately -0.9 $\log_{10}(PGA_{obseved})$ results from data logger equipment malfunctioning